\documentclass[twocolumn,preprintnumbers,amsmath,amssymbm,prl]{revtex4}
\usepackage{epsfig}
\usepackage{graphicx}

\begin{document}
\title{Comment on arXiv:1810.00886: ``Strong Cosmic Censorship: the nonlinear
story"}
\author{Shahar Hod}
\affiliation{The Ruppin Academic Center, Emeq Hefer 40250, Israel}
\affiliation{ }
\affiliation{The Hadassah Institute, Jerusalem 91010, Israel}
\date{\today}

\begin{abstract}
\ \ \ A very interesting work that has recently appeared in the
physics literature [arXiv:1810.00886] claims that the strong cosmic
censorship conjecture, a fundamental cornerstone of black-hole
physics, may be violated in asymptotically de Sitter charged
spacetimes. In this short Comment I would like to point out that
this conclusion cannot be self-consistently deduced from the
numerical simulations presented in arXiv:1810.00886.
\end{abstract}
\bigskip
\maketitle

The celebrated strong cosmic censorship conjecture (SCCC),
introduced by Penrose five decades ago \cite{Pen}, asserts that,
starting with spatially regular initial data, the dynamics of
self-gravitating field configurations will always produce globally
hyperbolic spacetimes. This conjecture implies, in particular, that
Cauchy horizons inside charged and spinning black holes should be
unstable to the infalling matter fields.

A very interesting work [arXiv:1810.00886] that has recently
appeared in the physics arXiv \cite{Car} argues that the SCCC may be
violated by charged Reissner-Nordstr\"om black holes in de Sitter
spacetimes. In particular, the authors of \cite{Car} claim that
``...our results show that SCC is not enforced by the field
equations." The intriguing claims made in \cite{Car} are based on
numerical studies of the dynamics of {\it neutral} (rather than {\it
charged}) matter fields in the non-asymptotically flat {\it charged}
spacetimes.

The main goal of the present short Comment is to point out that the
recent intriguing claims presented in \cite{Car} for a possible
violation of the SCCC in charged black-hole spacetimes are based on
incomplete matter models.

In particular, most physicists would agree that self-gravitating
{\it charged} matter fields are an inevitable part of any physically
realistic scenario of {\it charged} gravitational collapse that
starts with spatially regular initial conditions and which
eventually produces a charged black-hole spacetime. Thus, any
physically self-consistent test of the SCCC in charged spacetimes
should start with a {\it spatially regular} horizonless spacetime
which is coupled to spatially regular (smooth) {\it charged} matter
fields. One should then use the nonlinearly coupled
Einstein-charged-matter field equations in order to follow the
evolution of the charged matter fields (and the curved spacetime
itself) all the way to the formation of a charged black hole.

The interesting numerical model presented in \cite{Car} does {\it
not} meet the above stated physical requirements for a
self-consistent test of the SCCC in charged black-hole spacetimes
\cite{Notescr}. In particular, the presence of {\it charged} matter
fields as an inevitable part of the collapsed matter configurations
in {\it charged} black-hole spacetimes has been inconsistently
ignored in \cite{Car}. For this reason we believe that the
intriguing claim ``...our results show that SCC is not enforced by
the field equations" made in arXiv:1810.00886 \cite{Car} cannot be
self-consistently deduced from the numerical results presented in
this interesting work \cite{Notelin,Hodcom}.

\bigskip
\noindent
{\bf ACKNOWLEDGMENTS}
\bigskip

This research is supported by the Carmel Science Foundation. I would
like to thank Yael Oren, Arbel M. Ongo, Ayelet B. Lata, and Alona B.
Tea for helpful discussions.



\begin{thebibliography}{99}

\bibitem{Pen} S. W. Hawking and R. Penrose, Proc. R. Soc. Lond. A {\bf 314}, 529 (1970);
R. Penrose, Riv. Nuovo Cimento I {\bf 1}, 252 (1969); R. Penrose in
{\it General Relativity, an Einstein Centenary Survey}, eds. S.W.
Hawking and W. Israel (Cambridge University Press, 1979).

\bibitem{Car} R. Luna, M. Zilh\~ao, V.
Cardoso, J. L. Costa, and J. Nat\'ario, arXiv:1810.00886.

\bibitem{Notescr} This self-consistent physical procedure has been explicitly done
in asymptotically flat spacetimes two decades ago \cite{HodPir}.

\bibitem{HodPir} S. Hod and T. Piran, Phys. Rev. Lett. {\bf 81}, 1554
(1998) [arXiv:gr-qc/9803004]; S. Hod and T. Piran, Gen. Rel. Grav.
{\bf 30}, 1555 (1998) [arXiv:gr-qc/9902008. This essay received the
Second Prize from the Gravity Research Foundation 1998].

\bibitem{Notelin} It is worth mentioning that, using analytical techniques at the
linearized level, I have recently provided compelling evidence
\cite{Hodcom} that, by taking into account the inevitable presence
of {\it charged} matter fields in any self-consistent dynamical
model of {\it charged} black-hole formation, the strong cosmic
censorship conjecture is respected in non-asymptotically flat
charged black-hole spacetimes.

\bibitem{Hodcom} S. Hod, arXiv:1801.07261; S. Hod, Phys. Lett. B {\bf 780}, 221
(2018) [arXiv:1803.05443].

\end{thebibliography}
\end{document}